\documentstyle[prd,aps,preprint,epsf]{revtex}
\begin{document}
\thispagestyle{empty}
\draft
\hoffset=-10pt
\preprint{MSUHEP\,61022}{DOE-ER40757-088}{UTEXAS-HEP-96-19}
\title{Radiative Higgs boson decays   
\mbox{\boldmath $H\rightarrow f$}$\bar{\mbox{\boldmath $f$}}$\boldmath$\gamma$}

\author{Ali Abbasabadi$^1$, David Bowser-Chao$^2$, Duane A. Dicus$^3$ 
and Wayne W. Repko$^2$}
\address{$^1$Department of Physical Sciences \\
Ferris State University, Big Rapids, Michigan 49307}
\address{$^2$Department of Physics and Astronomy \\
Michigan State University, East Lansing, Michigan 48824}
\address{$^3$Center for Particle Physics and Department of Physics\\
University of Texas, Austin, Texas 78712}

\date{\today}
\maketitle 
\begin{abstract}
Higgs boson radiative decays of the form $H\rightarrow f\bar{f}\gamma$ 
are calculated in the Standard Model  using the complete one-loop expressions 
for the decay amplitudes. Contributions to the radiative width from leptons and
light quarks are given. We also present $e\bar{e}$ invariant mass 
distributions for $H\rightarrow e\bar{e}\gamma$,  which illustrate the 
importance of the photon pole contribution and the effects of the box diagrams. 
\end{abstract}
\newpage
\section{Introduction}

    Discussions of  searches for intermediate mass Higgs bosons usually  
concentrate on the decay $H\rightarrow \gamma\gamma$ as the discovery 
mode \cite{hhg}.  Other modes considered 
include $H\rightarrow Z\gamma$ and $H\rightarrow b\bar{b}$. The importance of 
radiative processes led us to consider the class of decays $H\rightarrow f
\bar{f} \gamma$, where $f$ is a light fermion. The dominant contributions to 
these decays occur at the one loop level, and their calculation is related to 
that of the process $e\bar{e} \rightarrow H\gamma$, which we recently 
completed \cite{ab-cdr,ddhr}. 

Typical results for Higgs boson $f\bar{f}\gamma$ decay appear in the
calculation of $\Gamma(H\rightarrow e\bar{e}\,\gamma)$, 
which, for $m_H\gtrsim 100\,$ GeV, receives a large contribution from the $Z$ 
pole. Additionally, our calculations show that the photon pole makes a 
substantial
correction to the estimate obtained by simply multiplying the width for $H
\rightarrow Z\,\gamma$ by the branching ratio $B(Z\rightarrow e\,\bar{e})$. 
This feature is common to all light fermions, and we present results for all
decays of the type $H \rightarrow f\bar{f}\gamma$. 

In the next section, we present expressions for the decay amplitudes, the decay
matrix element and results for the fermion invariant mass distributions and for
the widths. This is followed by a discussion. Complete expressions for the
various amplitudes are given in the appendices.

\section{Calculation of Higgs decay widths}

\begin{figure}
\hspace*{2.16in}
\epsfysize=2.0in \epsfbox{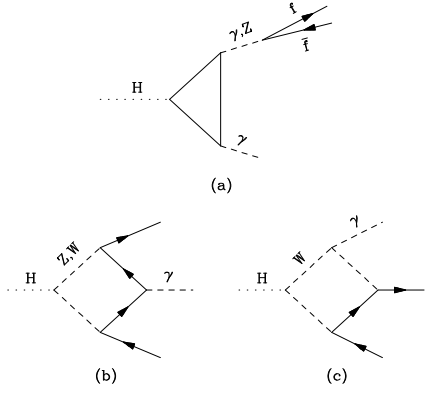}
\vspace{.10in}
\caption{The diagrams for $H\protect\rightarrow f\protect\bar{f}\gamma$ are
shown.}
\label{fig1}
\end{figure}

Contributions to the decay amplitudes arise from the diagrams illustrated 
\cite{tri} in
Fig\,(\ref{fig1}). They are of two basic types: ({\it a}) pole diagrams in which
the $f\bar{f}$ emerge from a virtual gauge boson; and ({\it b}) box diagrams
containing virtual gauge bosons and fermions in the loop. The diagrams are
evaluated using the non--linear gauges discussed in Ref.\cite{ab-cdr}. In these
gauges, the collection of diagrams consists of four separately gauge invariant 
contributions, the photon pole, the $Z$ pole, $Z$ boxes and $W$ boxes. The
amplitudes for these contributions are
\begin{eqnarray}
{\cal M}^{\gamma}_{\rm pole} & = & -\frac{\alpha^2m_W}{\sin\!\theta_W}\,
\bar{u}(p_1)\gamma_{\mu}v(p_2)\Bigl(\frac{
\delta_{\mu\nu}k\!\cdot\!(p_1 + p_2) - k_{\mu}(p_1 + p_2)_{\nu}}{m_{f\bar{f}}^2
}\Bigr)\hat{\epsilon}_{\nu}(k){\cal A}_{\gamma}(m_{f\bar{f}}^2)\,, \\ [4pt]
{\cal M}^{Z}_{\rm pole} & = & -\frac{\alpha^2m_W}{\sin^3\!\theta_W}\,\bar{u}
(p_1)\gamma_{\mu}(v_f + \gamma_5)v(p_2)\Bigl(\frac{\delta_{\mu\nu}k\!\cdot
\!(p_1 + p_2) - k_{\mu}(p_1 + p_2)_{\nu}}{(m_{f\bar{f}}^2 - m_Z^2) + 
im_Z\Gamma_Z}\Bigr)\hat{\epsilon}_{\nu}(k){\cal A}_{Z}(m_{f\bar{f}}^2)\,, 
\\ [4pt]
{\cal M}_{\rm box}^Z & = &\,\frac{\alpha^2m_Z}{4\sin^3\!\theta_W\cos^3\!
\theta_W}\,\bar{u}(p_1)\gamma_{\mu}(v_f + \gamma_5)^2v(p_2)
\Bigl\{\Bigl[\delta_{\mu\nu}k\!\cdot\!p_1 - k_{\mu}(p_1)_{\nu}
\Bigr]{\cal B}_Z(m_{f\bar{f}}^2,m_{f\gamma}^2,m_{\bar{f}\gamma}^2)
\nonumber \\
&   &+ \Bigl[\delta_{\mu\nu}k\!\cdot\!p_2 - k_{\mu}(p_2)_{\nu}\Bigr]
{\cal B}_Z(m_{f\bar{f}}^2,m_{\bar{f}\gamma}^2,m_{f\gamma}^2)\Bigr\}
\hat{\epsilon}_{\nu}(k)\;, \\ [4pt]
{\cal M}_{\rm box}^{W} & = &-\frac{\alpha^2m_W}{2\sin^3\!\theta_W}\,
\bar{u}(p_1)\gamma_{\mu}(1 + \gamma_5)^2v(p_2) 
\Bigl\{\Bigl[\delta_{\mu\nu}k\!\cdot\!p_1 - k_{\mu}(p_1)_{\nu}
\Bigr]{\cal B}_W(m_{f\bar{f}}^2,m_{f\gamma}^2,m_{\bar{f}\gamma}^2)\nonumber \\
&   &+ \Bigl[\delta_{\mu\nu}k\!\cdot\!p_2 - k_{\mu}(p_2)_{\nu}\Bigr]
{\cal B}_W(m_{f\bar{f}}^2,m_{\bar{f}\gamma}^2,m_{f\gamma}^2)\Bigr\}
\hat{\epsilon}_{\nu}(k)\;,
\end{eqnarray}
where $m_{f\bar{f}}^2 = -(p_1 + p_2)^2,\,m_{f\gamma}^2 = -(k + p_1)^2\,$ and 
$m_{\bar{f}\gamma}^2 = -(k + p_2)^2$. Here,
$v_f$ denotes the $f\bar{f}Z$ vector coupling constant, $v_f = 1 -
4|e_f|\sin^2\!\theta_W$, and $e_f$ is the fermion charge in units of the proton
charge. To calculate the
invariant amplitudes ${\cal B}_Z(m_{f\bar{f}}^2,m_{f\gamma}^2,m_{\bar{f}\gamma}
^2)$ and ${\cal B}_W(m_{f\bar{f}}^2,m_{f\gamma}^2,m_{\bar{f}\gamma}^2)$, we 
use the approach of Ref.\,\cite{ab-cdr}. Here, too, we find a logarithmic
dependence on the fermion mass at intermediate stages of the calculation, but
this dependence cancels enabling us to take the limit of zero fermion mass.
Explicit expressions for the invariant amplitudes ${\cal A}_{\gamma}$, ${\cal
A}_Z$, ${\cal B}_Z$ and ${\cal B}_W$ are given in the Appendices.

The invariant mass distribution $d\Gamma/dm_{f\bar{f}}^2$ is given by
\begin{equation}\label{dgam}
\frac{d\Gamma(H\rightarrow f\bar{f}\gamma)}{dm_{f\bar{f}}^2} =
\frac{1}{256\pi^3}\frac{1}{m_H^3}
\int_{(m_{\bar{f}\gamma}^2)_{\rm min}}^{(m_{\bar{f}\gamma}^2)_{\rm max}} 
dm_{\bar{f}\gamma}^2\,\sum_{\rm spin}|{\cal M}|^2\,,
\end{equation}
with the amplitude ${\cal M}$ given by the sum of Eqs.\,(1)--(4). For an
$f\bar{f}\gamma$ final state, the limits on the $dm_{\bar{f}\gamma}^2$ 
integration are
\begin{eqnarray}
(m_{\bar{f}\gamma}^2)_{\rm min} & = &\;m_f^2 + 
\case{1}{2}(m_H^2 - m_{f\bar{f}}^2)\left(1 - \sqrt{1 - \frac{\displaystyle 
4m_f^2}{\displaystyle m_{f\bar{f}}^2}}\;\;\right)\,, \\ [4pt]
(m_{\bar{f}\gamma}^2)_{\rm max} & = &\;m_f^2 + 
\case{1}{2}(m_H^2 - m_{f\bar{f}}^2)\left(1 + \sqrt{1 - 
\frac{\displaystyle 4m_f^2}{\displaystyle m_{f\bar{f}}^2}}\;\;\right)\,.
\end{eqnarray}
The fermion mass is retained in the phase space integration since, as shown 
below, there is a $(m_{f\bar{f}}^2)^{-1}$ factor associated with the photon 
pole. Explicitly, $\sum_{\rm spin}|{\cal M}|^2$ is
\begin{eqnarray}
\sum_{\rm spin}|{\cal M}|^2 & = &\;\frac{\alpha^4\,m_W^2\,m_{f\bar{f}}^2}
{16\sin^6\!\theta_W\,\cos^8\!\theta_W}
\biggl\{((m_{f\gamma}^2)^2 + (m_{\bar{f}\gamma}^2)^2)\Bigl[|\tilde{\cal A}_
{\gamma}|^2 + 
2v_f{\rm Re}(\tilde{\cal A}_{\gamma}\tilde{\cal A}_Z^*) 
\nonumber \\
&  &\;+ (1 + v_f^2)|\tilde{\cal A}_Z|^2\Bigr]
+ (m_{f\gamma}^2)^2\biggl[2(1 + v_f^2){\rm Re}(\tilde{\cal A}_{\gamma}
\tilde{\cal B}_Z^*) + 4{\rm Re}(\tilde{\cal A}_{\gamma}\tilde{\cal B}_W^*) 
\nonumber \\ [4pt] 
&  &\;+ 2(v_f^3 + 3v_f){\rm Re}(\tilde{\cal A}_Z\tilde{\cal B}_Z^*)
+ 4(1 + v_e){\rm Re}(\tilde{\cal A}_Z\tilde{\cal B}_W^*)
+ (1 + 6v_f^2 + v_f^4)|\tilde{\cal B}_Z|^2 \nonumber \\ [4pt]
&  &\;+ 4(1 + v_f)^2{\rm Re}(\tilde{\cal B}_Z\tilde{\cal B}_W^*)
+ 8|\tilde{\cal B}_W|^2\biggr] + 
(m_{\bar{f}\gamma}^2)^2\biggl[m_{f\gamma}^2\leftrightarrow m_{\bar{f}\gamma}^2
\biggr]\biggr\}\;.
\end{eqnarray}
Here, $\tilde{\cal A}_{\gamma}$, $\tilde{\cal A}_Z$, $\tilde{\cal B}_Z$ and 
$\tilde{B}_W$ are
\begin{eqnarray}
\tilde{\cal A}_{\gamma} & = &\,4\sin^2\!\theta_W\!\cos^4\theta_W\frac{{\cal A}_
{\gamma}(m_{f\bar{f}}^2)}{m_{f\bar{f}}^2}\,, \\ [4pt]
\tilde{\cal A}_Z   & = &\,4\cos^4\!\theta_W\frac{{\cal A}_Z(m_{f\bar{f}}^2)}
{(m_{f\bar{f}}^2 - m_Z^2) +im_Z\Gamma_Z}\,, \\ [4pt]
\tilde{\cal B}_Z   & = &\,-{\cal B}_Z(m_{f\bar{f}}^2,m_{f\gamma}^2,
m_{\bar{f}\gamma}^2)\,, \\ [4pt]
\tilde{\cal B}_W   & = &\,2\cos^4\!\theta_W\,{\cal B}_W(m_{f\bar{f}}^2,
m_{f\gamma}^2,m_{\bar{f}\gamma}^2)\,.
\end{eqnarray}
Using our results for the invariant amplitudes, the $dm_{\bar{f}\gamma}^2$
integration can be performed numerically to obtain the invariant mass
distribution.

The invariant mass distribution $d\Gamma(H\rightarrow e\bar{e}\gamma)
/dm_{e\bar{e}}$ is illustrated in Fig.\,(\ref{fig2}). The striking feature of
these distributions is the large peak at small $m_{e\bar{e}}^2$ due to the
photon pole. There is no singularity in the physical region since
$m_{e\bar{e}}^2\geq 4m_e^2$. In fact, as can be seen from Eqs.\,(6) and (7), the
$dm_{\bar{e}\gamma}^2$ integral in Eq.\,(\ref{dgam}) vanishes when 
$m_{e\bar{e}}^2 =
4m_e^2$. Nevertheless, the residual effect of the photon pole is sufficient to
contribute $\sim$ 10-20\% of the events in the distribution. It is also evident
that the box diagrams make only a small contribution. Curiously, the main effect
of the box diagrams is to smooth the distribution by cancelling the kinks in the
pole contributions at the $WW$ threshold. The invariant mass distribution for 
the remaining lepton channel $H\rightarrow \nu\bar{\nu}\gamma$, which has no 
contribution from the photon pole, is illustrated in Fig.\,(\ref{fig3}).

The various partial widths can be obtained by integrating Eq.\,(\ref{dgam}).
This results in the contributions illustrated in Fig.\,(\ref{fig4}). Also shown
in the lepton panel of Fig.\,(\ref{fig4}) is the contribution from the $Z$ pole.
The figure clearly shows that the widths are enhanced significantly in the
complete calculation, even in the case of neutrino decays. For
$m_H\gtrsim 160$ GeV, the up--type quark contributions are basically the same.
This is also true for the down--type quarks.

\section{Discussion}

\begin{figure}
\epsfysize=4.5in \epsfbox{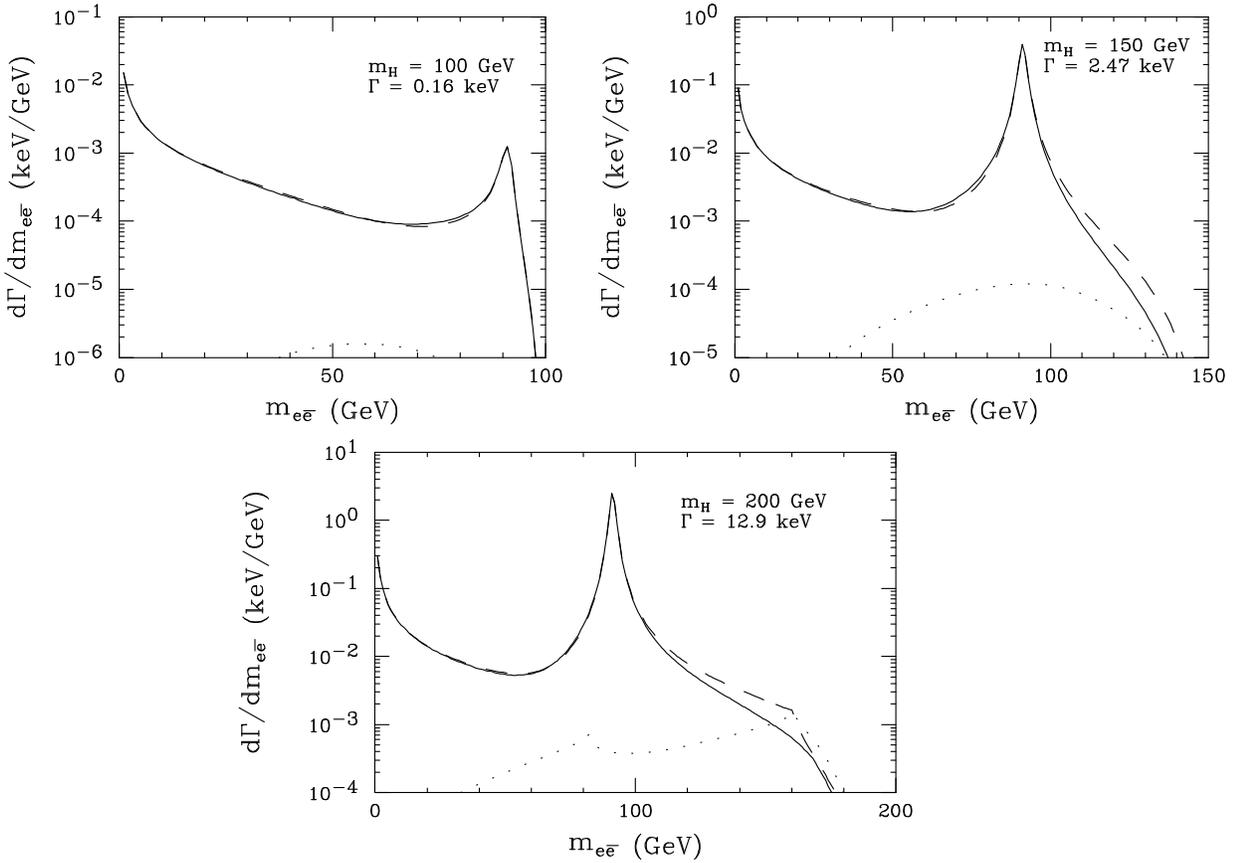}
\vspace{.10in}
\caption{The invariant mass distribution of the decay mode
$H\protect\rightarrow e\bar{e}\gamma$ for several Higgs masses is shown. The 
solid line is the full calculation, the dashed line is the pole contribution 
and the dotted line is the box contribution.}
\label{fig2}
\end{figure}

\begin{figure}
\hspace{1.0in}\epsfysize=2.5in \epsfbox{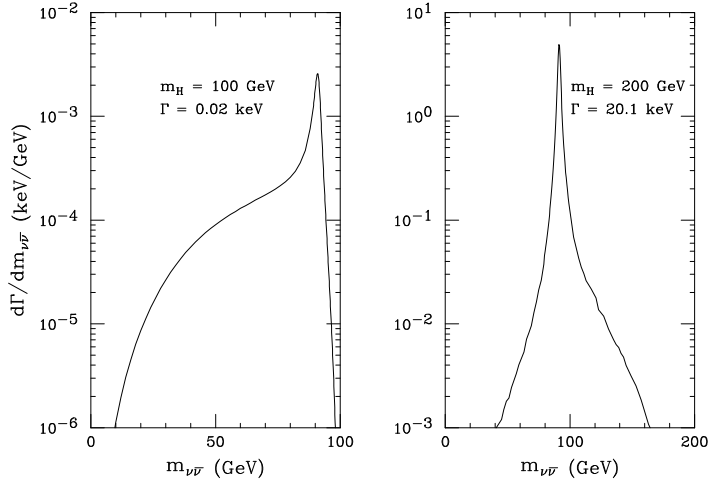}
\vspace{.10in}
\caption{The invariant mass distribution of the decay mode
$H\protect\rightarrow \nu\bar{\nu}\gamma$ is shown for two Higgs masses.} 
\label{fig3}
\end{figure}

\begin{figure}
\hspace{1.0in}\epsfysize=2.5in \epsfbox{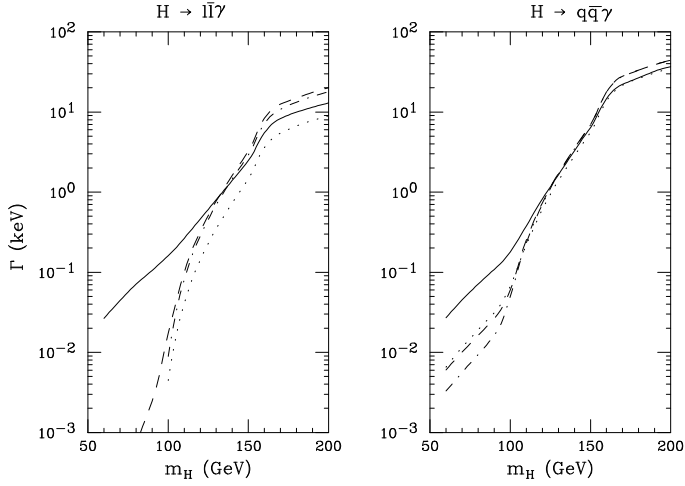}
\vspace{.10in}
\caption{The partial widths for Higgs decay into a lepton pair (left) and quark
pair (right) are plotted as a function of the Higgs mass. In the left panel, the
solid line is the partial width for a charged lepton and the dashed line is the
partial width for its neutrino. The dotted line is the $Z$ pole approximation
for the charged lepton partial width and the dot--dashed line is the 
pole approximation to the neutrino 
partial width. In the right panel, the solid line corresponds
to the up quark, the dashed line to the down quark, the dot-dashed line to the
strange quark and the dotted line to the charm quark.}
\label{fig4}
\end{figure}

As can be seen from Fig.\,(\ref{fig2}), the invariant mass distributions are
basically determined by the photon and $Z$ pole contributions. The box diagrams
make corrections to the high mass side of the distribution where they are of the
same order as the pole terms.
This being the case, it is possible to obtain a simplified expression for
$d\Gamma/dm_{f\bar{f}}^2$ by retaining only the ${\cal A}_{\gamma}$ and ${\cal
A}_Z$ terms in Eq.\,(\ref{dgam}). After performing the $dm_{\bar{f}\gamma}^2$
integration, one finds
\begin{eqnarray}
\frac{d\Gamma}{dm_{f\bar{f}}^2} & = &
\frac{\alpha^4\,m_W^2}{(8\pi)^3\sin^6\!\theta_Wm_H^3}
\Biggl[\sin^4\!\theta_W\frac{|{\cal A}_{\gamma}(m_{f\bar{f}}^2)|^2}
{m_{f\bar{f}}^2} + 2\sin^2\!\theta_W\,v_f{\rm Re}\biggl(\frac{{\cal A}_{\gamma}
(m_{f\bar{f}}^2){\cal A}_{Z}^*(m_{f\bar{f}}^2)}{(m_{f\bar{f}}^2 - m_Z^2) - 
im_Z\Gamma_Z}\biggr) \nonumber \\
&   &+ \frac{(1 + v_f^2)\,m_{f\bar{f}}^2|{\cal A}_{Z}(m_{f\bar{f}}^2)|^2}
{(m_{f\bar{f}}^2 - m_Z^2)^2 + m_Z^2\Gamma_Z^2}\Biggr](m_H^2 - m_{f\bar{f}}^2)
\sqrt{1 - \frac{\displaystyle 4m_f^2}{\displaystyle m_{f\bar{f}}^2}}\Biggl
[\biggl(m_H^2 + 2m_f^2 - m_{f\bar{f}}^2\biggr)^2 \nonumber \\
&   &+\;\case{1}{3}
\biggl(m_H^2 - m_{f\bar{f}}^2\biggr)^2\biggl(1 - \frac{4m_f^2}{m_{f\bar{f}}^2}
\biggr)\Biggr]\,.
\end{eqnarray}
This expression reproduces the dashed lines in Fig.\,(\ref{fig2}).

\begin{figure}
\hspace*{1.94in}
\epsfysize=2.4in \epsfbox{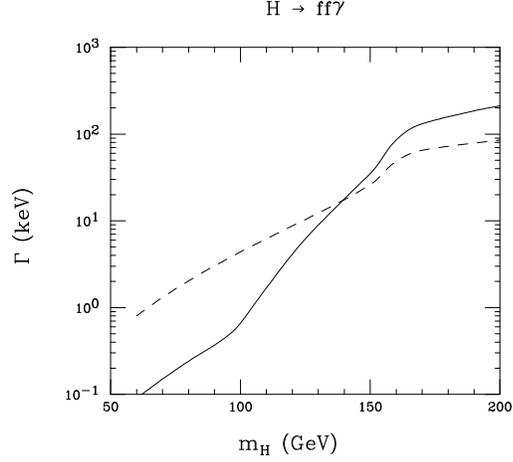}
\vspace{.10in}
\caption{The partial width $\Gamma(H\protect\rightarrow f\bar{f}\gamma)$,
obtained by summing the neutrino, electron, muon, up quark, down quark and 
strange quark contributions is shown as the solid line. For comparison, the 
dashed line is the partial width $\Gamma(H\protect\rightarrow\gamma\gamma)$.}
\label{fig5}
\end{figure}

The total width for $H\rightarrow f\bar{f}\gamma$ is shown in Fig.\,
(\ref{fig5}), where the neutrino, electron, muon, up quark, down quark and
strange quark contributions are added. The dashed line in this figure
corresponds to $H\rightarrow\gamma\gamma$, and it can be seen that the
$f\bar{f}\gamma$ width exceeds the $\gamma\gamma$ width for $m_H\gtrsim 140$
GeV.

\acknowledgements
This research was supported in part by the National Science Foundation under
grant PHY-93-07980 and by the United States Department of Energy under Contract
No. DE-FG013-93ER40757.

\appendix\section{Pole contributions}

The amplitudes ${\cal A}_\gamma(m_{f\bar{f}}^2)$ and ${\cal A}_Z
(m_{f\bar{f}}^2)$ can be expressed in terms of two scalar functions
$C_0(m_{f\bar{f}}^2,m_H^2,m^2)$ and $C_{23}(m_{f\bar{f}}^2,m_H^2,m^2)$, which
occur in the Passarino-Veltman decomposition \cite{pv} of the loop integrals, 
as 
\begin{eqnarray}
{\cal A}_{\gamma}(m_{f\bar{f}}^2) & = & -e_f\Bigl\{4(6 + \frac{m_H^2}{m_W^2})
C_{23}(m_{f\bar{f}}^2,m_H^2,m_W^2) - 16C_0(m_{f\bar{f}}^2,m_H^2,m_W^2) 
\nonumber \\
&  &\;-\,\frac{16}{3}\frac{m_t^2}{m_W^2}\Bigl(4C_{23}(m_{f\bar{f}}^2,m_H^2,
m_t^2) -  C_0(m_{f\bar{f}}^2,m_H^2,m_t^2)\Bigr)\Bigr\}\;,\\ [6pt]
{\cal A}_Z(m_{f\bar{f}}^2) & = &\; -2I_3\Bigl\{\Bigl(5 - \tan^2\theta_W
 + \frac{m_H^2}{2m_W^2}(1 - \tan^2\theta_W)\Bigr)
C_{23}(m_{f\bar{f}}^2,m_H^2,m_W^2) \nonumber \\
&  &\;+\, \Bigl(\tan^2\theta_W - 3\Bigr)
C_0(m_{f\bar{f}}^2,m_H^2,m_W^2) -\,\frac{1}{2}\frac{m^2_t}{m^2_W}\frac{1 - 
(8/3)\sin^2\theta_W}{\cos^2\!\theta_W}
\nonumber \\
&  &\;\times\Bigl(4C_{23}(m_{f\bar{f}}^2,m_H^2,m_t^2) - 
C_0(m_{f\bar{f}}^2,m_H^2,m_t^2)\Bigr)\Bigr\}\;,
\end{eqnarray}
with $I_3$ denoting the third component of the {\em external} fermion weak 
isospin and $m_t$ being the top quark mass. 

    The evaluation of $C_0(m_{f\bar{f}}^2,m_H^2,m^2)$ is straightforward, 
yielding
\begin{equation}
C_0(m_{f\bar{f}}^2,m_H^2,m^2)  =\;\frac{1}{(m_{f\bar{f}}^2 - m_H^2)}
\Bigl(C(\frac{m_H^2}{m^2}) - C(\frac{m_{f\bar{f}}^2}{m^2})\Bigr)\;,
\end{equation}
where
\begin{eqnarray}
C(\beta) & = &\;\int_0^1\frac{dx}{x}\ln\Bigl(1 - \beta x(1 - x) -
i\varepsilon\Bigr) \\
         & = &\;\left\{
\begin{array}{lll}
-2\Bigl(\sin^{-1}(\sqrt{\frac{\displaystyle\beta}{\displaystyle 4}}\,)\Bigr)^2 
 &  & 0\leq\beta\leq 4 \\
2\Bigl(\cosh^{-1}(\sqrt{\frac{\displaystyle\beta}{\displaystyle 4}}\,)\Bigr)^2 
- \frac{\displaystyle\pi^2}{\displaystyle2} - 2i\pi\cosh^{-1}(\sqrt{
\frac{\displaystyle\beta}{\displaystyle 4}}) &  & \beta\geq 4
\end{array}
\right.\,.
\end{eqnarray}

The scalar function $C_{23}(m_{f\bar{f}}^2,m_H^2,m^2)$ is expressible as
\begin{eqnarray}
C_{23}(m_{f\bar{f}}^2,m_H^2,m^2) & = &\;\frac{1}{2}\frac{1}{(m_{f\bar{f}}^2 - 
m_H^2)}\Bigl[1 + \frac{m_{f\bar{f}}^2}{(m_{f\bar{f}}^2 - m_H^2)}
\Bigl(B(\frac{m_H^2}{m^2}) - B(\frac{m_{f\bar{f}}^2}{m^2})\Bigr) \nonumber \\
&   &\;-\frac{2m^2}{(m_{f\bar{f}}^2 - m_H^2)}\Bigl(C(\frac{m_H^2}{m^2}) - 
C(\frac{m_{f\bar{f}}^2}{m^2})\Bigr)\Bigr]\;,
\end{eqnarray}
where the function $B(\beta)$ is 
\begin{eqnarray}
B(\beta) & = & \int_0^1dx\ln\bigl(1 - \beta x(1 - x) - i\varepsilon\Bigr) \\
         & = & \left\{
\begin{array}{lll}
2\Bigl[\sqrt{\frac{\displaystyle 4 - \beta}{\displaystyle\beta}}\sin^{-1}
(\sqrt{\frac{\displaystyle\beta}{\displaystyle 4}}\,) - 1\Bigr] &  & 
0\leq\beta\leq 4 \\
2\Bigl[\sqrt{\frac{\displaystyle\beta - 4}{\displaystyle\beta}}\cosh^{-1}
(\sqrt{\frac{\displaystyle\beta}{\displaystyle 4}}\,) - 1 -
\frac{\displaystyle i\pi}{\displaystyle 2}\sqrt{\frac{\displaystyle\beta - 4}
{\displaystyle\beta}}\,\Bigr] &  & \beta\geq 4
\end{array}
\right.\,.
\end{eqnarray}
\section{`${\bf Z}$' Box Contribution}
The `$Z$' box diagrams are illustrated in Fig.\,\ref{fig1}(b), where the
internal gauge boson can be either a $Z$ or a $W$. In addition to these box
diagrams, there are two triangle diagrams which make the amplitude gauge
invariant. The entire contribution is given by the function ${\cal B}_0
(m_{f\bar{f}}^2,m_{f\gamma}^2,m_{\bar{f}\gamma}^2)$ which is related to the
function ${\cal B}_Z$ appearing in Eq.\,(3) as
\begin{equation}
{\cal B}_Z(m_{f\bar{f}}^2,m_{f\gamma}^2,m_{\bar{f}\gamma}^2) = -e_f\,{\cal
B}_0(m_{f\bar{f}}^2,m_{f\gamma}^2,m_{\bar{f}\gamma}^2)\,.
\end{equation}
In terms of the decomposition of Ref.\cite{pv}, this function is given by
\begin{eqnarray}\label{bz}
{\cal B}_0(m_{f\bar{f}}^2,m_{f\gamma}^2,m_{\bar{f}\gamma}^2) & = &
D_0(m_{f\bar{f}}^2,m_{f\gamma}^2,m_{\bar{f}\gamma}^2,m_H^2,m_f^2,m_Z^2) 
+ D_{11}(m_{f\bar{f}}^2,m_{f\gamma}^2,m_{\bar{f}\gamma}^2,m_H^2,m_f^2,
m_Z^2) \nonumber \\
&   &+ D_{12}(m_{f\bar{f}}^2,m_{f\gamma}^2,m_{\bar{f}\gamma}^2,m_H^2,m_f^2,
m_Z^2) \nonumber  \\
&   &+ D_{24}(m_{f\bar{f}}^2,m_{f\gamma}^2,m_{\bar{f}\gamma}^2,m_H^2,m_f^2,
m_Z^2). 
\end{eqnarray}
When the $D_{\alpha\beta}$ in Eq.\,(\ref{bz}) are expanded in terms of scalar
integrals, the expression for ${\cal B}_0(m_{f\bar{f}}^2,m_{f\gamma}^2,
m_{\bar{f}\gamma}^2)$ takes the form
\begin{eqnarray}\label{bzt} 
{\cal B}_0(m_{f\bar{f}}^2,m_{f\gamma}^2,m_{\bar{f}\gamma}^2) & = &
\frac{1}{2}\frac{1}{m_{f\bar{f}}^2\,m_{f\gamma}^2}\Biggl\{
(1 - \frac{m_Z^2}{m_{f\gamma}^2})\Bigl(m_Z^2(m_{f\gamma}^2 + m_{\bar{f}
\gamma}^2) - m_{f\gamma}^2m_{\bar{f}\gamma}^2\Bigr)D_0(1,2,3,4) \nonumber \\
&   &- (1 - \frac{m_Z^2}{m_{f\gamma}^2})\Bigl[m_{f\gamma}^2C_0(1,2,3) -
(m_{f\bar{f}}^2 + m_{\bar{f}\gamma}^2)C_0(1,3,4)\Bigr]\\ 
&   &-\Bigl[(1 - \frac{m_Z^2}{m_{f\gamma}^2})(m_{f\bar{f}}^2 - 
m_{f\gamma}^2) - 2m_Z^2\frac{m_{f\bar{f}}^2}{(m_{f\bar{f}}^2 + m_{f\gamma}^2)}
\Bigr]C_0(1,2,4) \nonumber \\
&   &- m_{\bar{f}\gamma}^2(1 - \frac{m_Z^2}{m_{f\gamma}^2})
C_0(2,3,4) + \frac{2m_{f\bar{f}}^2}{(m_{f\bar{f}}^2 + m_{f\gamma}^2)}
\Bigl[B_0(1,4) - B_0(2,4)\Bigr]\Biggr\}, \nonumber
\end{eqnarray}
where we have used the compact notation of Ref.\,\cite{pv} with $m_1 = m_4 =
m_Z$ and $m_2 = m_3 = m_f$, as illustrated in Fig.\,\ref{fig6}\,(a). 

Explicitly, we have
\begin{eqnarray}
D_0(1,2,3,4) & = & -\frac{1}{\tau}\Biggl\{
{\rm Sp}\left(\frac{m_{f\gamma}^2 - m_Z^2}{-m_Z^2}\right)
- {\rm Sp}\left(\frac{-m_Z^2}{m_{f\gamma}^2 - m_Z^2}\right)\nonumber \\
&   & 
+ {\rm Sp}\left(\frac{m_{\bar{f}\gamma}^2(m_{f\gamma}^2 - m_Z^2)}{\tau}\right)
- {\rm Sp}\left(\frac{-m_{\bar{f}\gamma}^2m_Z^2}{\tau}\right) \nonumber \\
&   & 
- {\rm Sp}\left(\frac{m_{f\gamma}^2 - m_Z^2}{m_{f\gamma}^2(1 - \beta_{Z+}) - 
m_Z^2}\right)
+ {\rm Sp}\left(\frac{-m_Z^2}{m_{f\gamma}^2(1 - \beta_{Z+}) - m_Z^2}\right)
\nonumber \\
&   &
- {\rm Sp}\left(\frac{m_{f\gamma}^2 - m_Z^2}{m_{f\gamma}^2(1 - \beta_{Z-}) - 
m_Z^2}\right)
+ {\rm Sp}\left(\frac{-m_Z^2}{m_{f\gamma}^2(1 - \beta_{Z-}) - m_Z^2}\right)
\nonumber \\
&   & 
+ {\rm Sp}\left(\frac{m_{f\gamma}^2 - m_Z^2}{m_Z^2\,m_{f\gamma}^2
(\delta_0 - \delta_2)}\right)  
- {\rm Sp}\left(\frac{- 1}{m_{f\gamma}^2(\delta_0 - \delta_2)}
\right) \nonumber \\
&   & 
+ {\rm Sp}\left(\frac{m_Z^2m_{f\gamma}^2(\delta_0 - \delta_2)}
{m_{f\gamma}^2 - m_Z^2}\right) 
- {\rm Sp}\left(\frac{m_{f\gamma}^2(\delta_0 - \delta_2)}{- 1}
\right)\nonumber \\
&   & 
+ \ln\left(\frac{m_f^2}{m_Z^2}\right)\ln\left|\frac{m_{f\gamma}^2 -
m_Z^2}{-m_Z^2}\right|
+ i\pi\Biggl[\theta(m_H^2 - 4m_Z^2)
\ln\left|\frac{\beta_{Z+} - \alpha_0}{\beta_{Z-} -
\alpha_0}\right|\nonumber \\
&   &
+ \theta(m_{f\gamma}^2 - m_Z^2)\Biggl(\ln\left|
\frac{m_{\bar{f}\gamma}^2(m_{f\gamma}^2 - m_Z^2)}{m_{f\gamma}^2m_Z^2}\right| 
+ \ln\left|\frac{\tau}{m_{f\gamma}^2 m_{\bar{f}\gamma}^2}\right|\nonumber \\
&   &
- \ln\left|m_f^2(\delta_0 -\delta_2)\right|\Biggr)\Biggr] 
+ m_{f\gamma}^2\leftrightarrow m_{\bar{f}\gamma}^2\Biggr\}\,,
\end{eqnarray}
where 
\begin{eqnarray}
\tau & = &m_{f\gamma}^2m_{\bar{f}\gamma}^2 - m_Z^2(m_{f\gamma}^2 + 
m_{\bar{f}\gamma}^2) \\
\alpha_0 & = & 1 - \frac{m_Z^2}{m_{f\gamma}^2} + m_f^2\delta_0 \\
\beta_{Z\pm} & = & \case{1}{2}\left(1 \pm \sqrt{1 - 4m_Z^2/m_H^2}\;\right) \,,
\end{eqnarray}
with
\begin{eqnarray}
\delta_0 & = & \frac{m_{f\bar{f}}^2}{\tau} \\
\delta_2 & = & \frac{-1}{(m_{f\gamma}^2 - m_Z^2)}\,.
\end{eqnarray}
The function Sp($y$) is defined as the real part of the
dilogarithm. For real $y > 1$, we have
\begin{equation}
{\rm Sp}(y) = {\rm Re} \left(-\int_0^1 \frac{dx}{x} \ln(1-yx+i\varepsilon)
\right)\,.
\end{equation}
When $y$ is complex, the $i\varepsilon$ may be ignored.

The required $C_0$ functions are
\begin{eqnarray}
C_0(1,2,3) 
& = & \frac{-1}{m_{f\gamma}^2}\Biggl\{
{\rm Sp}\left(\frac{m_Z^2 - m_{f\gamma}^2}{m_{f\gamma}^2}\right) 
+ {\rm Sp}\left(\frac{m_{f\gamma}^2}{m_Z^2 - m_{f\gamma}^2}\right) \nonumber \\
&   &
- {\rm Sp}\left(\frac{(m_Z^2 - m_{f\gamma}^2)^2}{m_Z^2m_{f\gamma}^2}\right)
- {\rm Sp}\left(\frac{m_Z^2m_{f\gamma}^2}{(m_Z^2 - m_{f\gamma}^2)^2}\right)
\nonumber \\
&   &
+ {\rm Sp}\left(\frac{m_Z^2}{m_Z^2 - m_{f\gamma}^2}\right) - \frac{\pi^2}{6}
+ \ln\left(\frac{m_f^2}{m_Z^2}\right)\ln\left|\frac{m_Z^2}{m_Z^2 -
m_{f\gamma}^2}\right|\nonumber \\
&   &
-i\pi\theta(m_{f\gamma}^2 - m_Z^2)\ln\left|
\frac{(m_Z^2 - m_{f\gamma}^2)^2}{m_f^2m_{f\gamma}^2}\right|\;\Biggr\}\,,\\ [8pt]
C_0(2,3,4)
& = & \frac{1}{m_{\bar{f}\gamma}^2}\Biggl\{
{\rm Sp}\left(\frac{m_{\bar{f}\gamma}^2 - m_Z^2}{m_Z^2}\right)
+ {\rm Sp}\left(\frac{m_Z^2}{m_{\bar{f}\gamma}^2 - m_Z^2}\right) \nonumber \\
&   &
+ {\rm Sp}\left(\frac{m_{\bar{f}\gamma}^2}{m_{\bar{f}\gamma}^2 - m_Z^2}\right)
+ \frac{\pi^2}{6} + \ln\left(\frac{m_f^2}{m_Z^2}\right)\ln\left|\frac{m_Z^2 -
m_{\bar{f}\gamma}^2}{m_Z^2}\right| \nonumber \\
&   &
+ i\pi\theta(m_{\bar{f}\gamma}^2 - m_Z^2)\ln\left|\frac{(m_Z^2 - m_{\bar{f}
\gamma}^2)^2}{m_f^2m_{\bar{f}\gamma}^2}\right|\;\Biggr\}\,, \\ [8pt]
C_0(1,2,4) 
& = &\frac{-1}{m_{\bar{f}\gamma}^2 - m_H^2}\Biggl\{
{\rm Sp}\left(\frac{\gamma_0}{\gamma_0 - \gamma_{1}}\right)
- {\rm Sp}\left(\frac{\gamma_0 - 1}{\gamma_0 - \gamma_{1}}\right)
- {\rm Sp}\left(\frac{\gamma_0}{\gamma_0 - \beta_{Z+}}\right)\nonumber \\
&   &
+ {\rm Sp}\left(\frac{\gamma_0 - 1}{\gamma_0 - \beta_{Z+}}\right)
- {\rm Sp}\left(\frac{\gamma_0}{\gamma_0 - \beta_{Z-}}\right)
+ {\rm Sp}\left(\frac{\gamma_0 - 1}{\gamma_0 - \beta_{Z-}}\right) \nonumber \\
&   &
- {\rm Sp}\left(\frac{\gamma_0 - 1}{\gamma_0}\right) + \frac{\pi^2}{6} \nonumber
\\
&   &
+ i\pi\Biggl[\theta(m_H^2 - 4m_Z^2)\ln\left|\frac{\beta_{Z+} - \gamma_0}
{\beta_{Z-} - \gamma_0}\right|
- \theta(m_{\bar{f}\gamma}^2 - m_Z^2)\ln\left|\frac{\gamma_1 -
\gamma_0}{-\gamma_0}\right|\;\Biggr]\;\Biggr\}\,,
\end{eqnarray}
with
\begin{eqnarray}
\gamma_0 & = & \frac{m_Z^2}{m_H^2 - m_{\bar{f}\gamma}^2}\\
\gamma_1 & = & \frac{m_{\bar{f}\gamma}^2 - m_Z^2}{m_{\bar{f}\gamma}^2}\,.
\end{eqnarray}
The expression for $C_0(1,3,4)$ is obtained from $C_0(1,2,4)$ by the 
interchange $m_{f\gamma}^2\leftrightarrow m_{\bar{f}\gamma}^2$.

Finally, the required $B_0$ functions are
\begin{eqnarray} \label{b0z}
B_0(1,4) & = &\;\Delta + 2
+ \beta_{Z+}\ln\left(\frac{\beta_{Z+} - 1 + i\varepsilon}{\beta_{Z+}}\right) 
+ \beta_{Z-}\ln\left(\frac{\beta_{Z-} - 1 - i\varepsilon}{\beta_{Z-}}\right)
\,, \\ [8pt]
B_0(2,4) & = &\;\Delta + 2 
- \left(1 - \frac{m_Z^2}{m_{\bar{f}\gamma}^2}\right)\Biggl[\ln\left|
1 - \frac{m_{\bar{f}\gamma}^2}{m_Z^2}\right|
- i\pi\theta(m_{\bar{f}\gamma}^2 - m_Z^2)\,\Biggr]\,,
\end{eqnarray}
where $\Delta$ is 
\begin{equation}\label{delta}
\Delta = \pi^{(n/2 -2)}\left(\frac{m_Z^2}{\mu^2}\right)^{(n/2 -2)}
\Gamma\left(2 - \frac{n}{2}\right)\,,
\end{equation}
and $\mu$ is introduced to preserve the dimensions of the regularized integrals.

    The cancellation of the $\ln(m_f^2)$ dependence in Eqs.\,(B4), (B11) and
(B12) can be checked by substituting the explicit expressions into Eq.\,(B3).

\section{${\bf W}$ Box Contribution}
\begin{figure}[h]
\hspace*{1.94in}
\epsfysize=2.4in \epsfbox{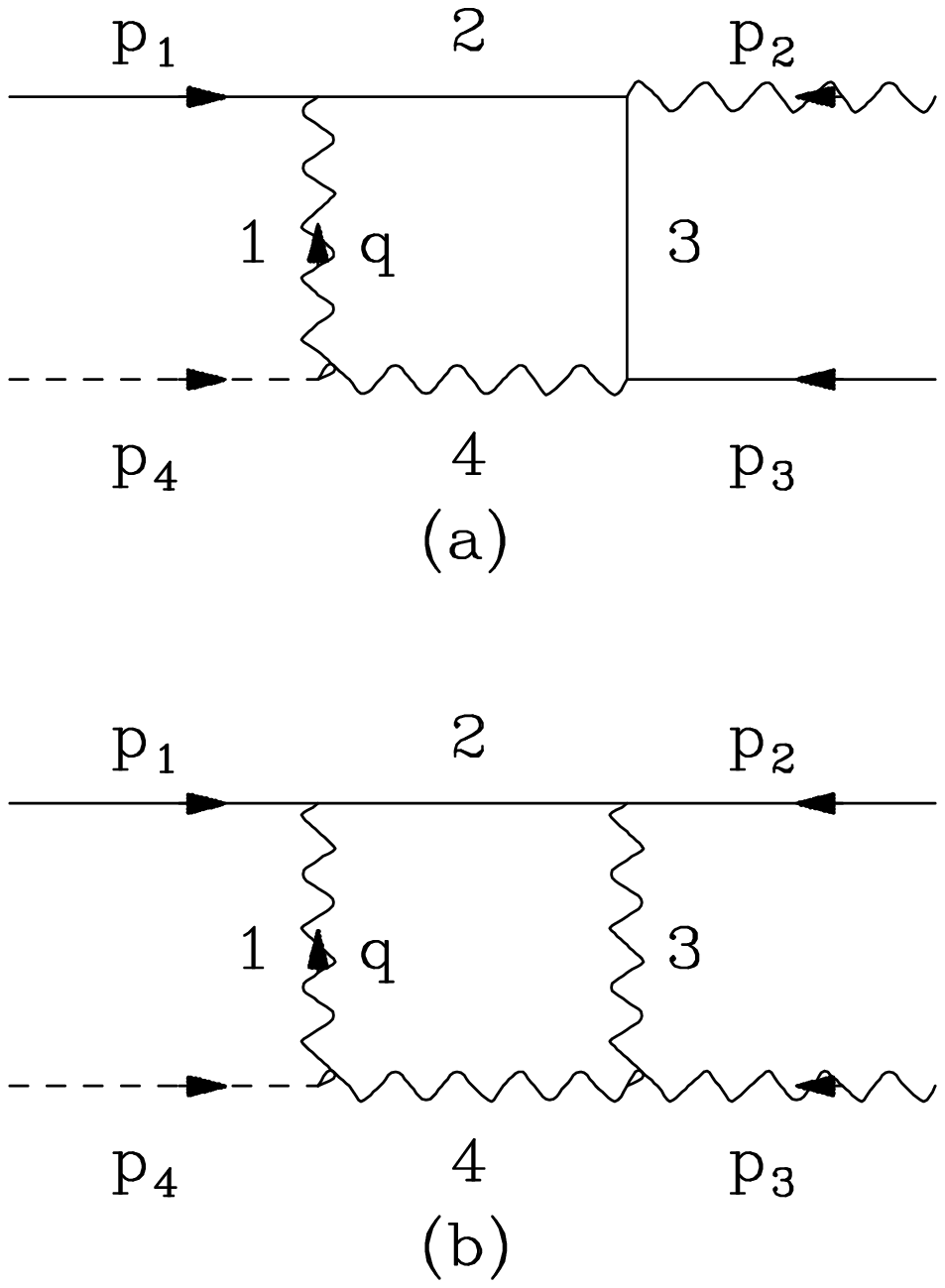}
\vspace{.10in}
\caption{The numbering schemes used for the computation of $D_0(1,2,3,4)$ in the
case of the $Z$ box (a) and the $W$ box (b) are shown.}
\label{fig6}
\end{figure}

The box diagrams with $W$'s in the loop are shown in Fig.\,1(b) and Fig.\,1(c).
The non-gauge-invariant portions of these diagrams are again canceled by 
triangle diagrams in which the Higgs boson decays into an $f\bar{f}$ pair 
through a $WWf$ triangle and the photon is emitted from one of the fermions. The
invariant amplitude ${\cal B}_W(m_{f\bar{f}}^2,m_{f\gamma}^2,
m_{\bar{f}\gamma}^2)$ appearing in Eq.\,(4) can be expressed as 
\begin{eqnarray}
{\cal B}_W(m_{f\bar{f}}^2,m_{f\gamma}^2,m_{\bar{f}\gamma}^2) & = &-2I_3
\Biggl({\cal B}_1(m_{f\bar{f}}^2,m_{f\gamma}^2,m_{\bar{f}\gamma}^2) + {\cal
B}_2(m_{f\bar{f}}^2,m_{\bar{f}\gamma}^2,m_{f\gamma}^2)\Biggr) \nonumber \\
&   &+\,e_i{\cal B}_0^{\prime}(m_{f\bar{f}}^2,m_{f\gamma}^2,
m_{\bar{f}\gamma}^2)\,,
\end{eqnarray}
where, as in Appendix A, $I_3$ is the third component of the external fermion
weak isospin. Here, $e_i$ is the charge of the {\em internal} fermion in units 
of the proton charge and the prime denotes the
replacement $m_Z\rightarrow m_W$ in Eq.\,(\ref{bz}). Notice that the arguments 
$m_{f\gamma}^2$ and $m_{\bar{f}\gamma}^2$ are interchanged in ${\cal B}_2$.
In terms of the $D_{\alpha\beta}$ of Ref.\cite{pv}, we have
\begin{eqnarray}
{\cal B}_1(m_{f\bar{f}}^2,m_{f\gamma}^2,m_{\bar{f}\gamma}^2) & = &
D_0(m_{f\bar{f}}^2,m_{f\gamma}^2,m_{\bar{f}\gamma}^2,m_H^2,m_f^2,m_W^2)
+ D_{11}(m_{f\bar{f}}^2,m_{f\gamma}^2,m_{\bar{f}\gamma}^2,m_H^2,m_f^2,m_W^2) 
\nonumber \\
&   &+D_{13}(m_{f\bar{f}}^2,m_{f\gamma}^2,m_{\bar{f}\gamma}^2,m_H^2,m_f^2,
m_W^2) \nonumber \\
&   &+D_{25}(m_{f\bar{f}}^2,m_{f\gamma}^2,m_{\bar{f}\gamma}^2,m_H^2,m_f^2,
m_W^2)\,, \\ [6pt]
{\cal B}_2(m_{f\bar{f}}^2,m_{f\gamma}^2,m_{\bar{f}\gamma}^2) & = &-
D_{12}(m_{f\bar{f}}^2,m_{f\gamma}^2,m_{\bar{f}\gamma}^2,m_H^2,m_f^2,m_W^2)
\nonumber \\
&   &+D_{13}(m_{f\bar{f}}^2,m_{f\gamma}^2,m_{\bar{f}\gamma}^2,m_H^2,m_f^2,
m_W^2) \nonumber \\
&   &+D_{26}(m_{f\bar{f}}^2,m_{f\gamma}^2,m_{\bar{f}\gamma}^2,m_H^2,m_f^2,
m_W^2)\,.
\end{eqnarray}
The decomposition into scalar functions takes the form
\begin{eqnarray}\label{a1}
{\cal B}_1(m_{f\bar{f}}^2,m_{f\gamma}^2,m_{\bar{f}\gamma}^2) & = &\frac{1}{2}
\frac{1}{m_{f\bar{f}}^2\,m_{f\gamma}^2}\Biggl\{\Biggl[\frac{(m_{f\bar{f}}^2 + 
m_{f\gamma}^2 - m_W^2)}{m_{f\gamma}^2}
\left(m_W^2(m_{f\gamma}^2 + m_{\bar{f}\gamma}^2) -
m_{f\bar{f}}^2\,m_{\bar{f}\gamma}^2\right)  \nonumber \\
&   &- 2m_W^2m_{f\bar{f}}^2\Biggr]D_0(1,2,3,4) 
 + (m_{f\bar{f}}^2 + m_{f\gamma}^2  - m_W^2)\Bigl[
-\frac{m_{f\bar{f}}^2}{m_{f\gamma}^2}C_0(1,2,3)  \nonumber \\
&   &+ \frac{(m_{f\bar{f}}^4 + 2m_{f\bar{f}}^2\,m_{f\gamma}^2 - m_{f\gamma}^4)}
{(m_{f\bar{f}}^2 + m_{f\gamma}^2)m_{f\gamma}^2}C_0(1,2,4)
+ \frac{(m_{f\gamma}^2 + m_{\bar{f}\gamma}^2)}{m_{f\gamma}^2}C_0(1,3,4) 
\nonumber \\
&   &- \frac{m_{\bar{f}\gamma}^2}{m_{f\gamma}^2}C_0(2,3,4)\Bigr] 
+ \frac{2m_{f\bar{f}}^2}{(m_{f\gamma}^2 +
m_{\bar{f}\gamma}^2)}\left[B_0(1,3) - B_0(1,4)\right]  \nonumber \\
&   &+ \frac{2m_{f\bar{f}}^2}{(m_{f\bar{f}}^2 + m_{f\gamma}^2)}
\left[B_0(2,4) - B_0(1,4)\right]\Biggr\}\,,\\ [8pt] 
\label{a2}
{\cal B}_2(m_{f\bar{f}}^2,m_{f\gamma}^2,m_{\bar{f}\gamma}^2) & = &\;\frac{1}{2}
\frac{1}{m_{f\bar{f}}^2\,m_{\bar{f}\gamma}^2}\Biggl\{
\frac{(m_{\bar{f}\gamma}^2 - m_W^2)}{m_{\bar{f}\gamma}^2}
\left(m_{f\bar{f}}^2\,m_{\bar{f}\gamma}^2 +
m_W^2(m_{f\gamma}^2 + m_{\bar{f}\gamma}^2)\right)D_0(1,2,3,4) 
  \nonumber \\
&   & + (m_{\bar{f}\gamma}^2 - m_W^2)\Bigl[\;
\frac{m_{f\bar{f}}^2}{m_{\bar{f}\gamma}^2}C_0(1,2,3) 
- \frac{(m_{f\bar{f}}^2 + m_{f\gamma}^2)}{m_{\bar{f}\gamma}^2}
C_0(1,2,4)   \nonumber \\
&    & + \frac{(m_{f\gamma}^2 + m_{\bar{f}\gamma}^2)}{m_{\bar{f}\gamma}^2
}C_0(1,3,4) - C_0(2,3,4)\Bigr]  \nonumber \\
&   &+ \frac{2m_{f\bar{f}}^2}{(m_{f\gamma}^2 + m_{\bar{f}\gamma}^2)}
\left[B_0(1,4) - B_0(1,3)\right]\Biggr\}\,.
\end{eqnarray}
The numbering is defined in Fig.\,\ref{fig6}\,(b), with $m_1 = m_3 = m_4 = 
m_W$ and $m_2 = m_i = 0$.

In this case, the scalar function $D_0(1,2,3,4)$ is 
\begin{eqnarray}
D_0(1,2,3,4) & = & \frac{-1}{m_{f\bar{f}}^2m_{\bar{f}\gamma}^2}\frac{1}
{(\beta_{0+} - \beta_{0-})}\Biggl\{
{\rm Sp}\left(\frac{\beta_{0+}}{\beta_{0+} - m_W^2/m_{\bar{f}\gamma}^2}\right)
- {\rm Sp}\left(\frac{\beta_{0+} - 1}{\beta_{0+} - m_W^2/m_{\bar{f}\gamma}^2}
\right) \nonumber \\
&   &
- {\rm Sp}\left(\frac{\beta_{0+}}{\beta_{0+} - \beta_{W+}}\right)
+ {\rm Sp}\left(\frac{\beta_{0+} - 1}{\beta_{0+} - \beta_{W+}}\right)
- {\rm Sp}\left(\frac{\beta_{0+}}{\beta_{0+} - \beta_{W-}}\right) \nonumber \\
&   &
+ {\rm Sp}\left(\frac{\beta_{0+} - 1}{\beta_{0+} - \beta_{W-}}\right)
+{\rm Sp}\left(\frac{\beta_{0+}}{\beta_{0+} - \gamma_+}\right)
- {\rm Sp}\left(\frac{\beta_{0+} - 1}{\beta_{0+} - \gamma_+}\right) \nonumber
\\
&   &
+{\rm Sp}\left(\frac{\beta_{0+}}{\beta_{0+} - \gamma_-}\right)
- {\rm Sp}\left(\frac{\beta_{0+} - 1}{\beta_{0+} - \gamma_-}\right) \nonumber
\\
&   & +i\pi\Biggl[-\theta(m_{\bar{f}\gamma}^2 - m_W^2)\ln\left|\frac{\beta_{0+}
- 1}{\beta_{0+} - m_W^2/m_{\bar{f}\gamma}^2}\right|
+ \theta(m_H^2 - 4m_W^2)\ln\left|\frac{\beta_{W+} - \beta_{0+}}{\beta_{W-} -
\beta_{0+}}\right| \nonumber \\
&   &
- \theta(m_{f\bar{f}}^2 - 4m_W^2)\ln\left|\frac{\gamma_+ - \beta_{0+}}
{\gamma_- - \beta_{0+}}\right|\;\Biggr] +\;\beta_{0+}\rightarrow\beta_{0-}
\Biggr\}\,,
\end{eqnarray}
with
\begin{eqnarray}
\beta_{0\pm} & = & \case{1}{2}\left(1 - \frac{m_W^2}{m_{f\bar{f}}^2} -
\frac{m_W^2}{m_{f\bar{f}}^2}\frac{m_{f\gamma}^2}{m_{\bar{f}\gamma}^2} \pm
\sqrt{\left(1 - \frac{m_W^2}{m_{f\bar{f}}^2} - \frac{m_W^2}{m_{f\bar{f}}^2}
\frac{m_{f\gamma}^2}{m_{\bar{f}\gamma}^2}\right)^2 +
4\frac{m_W^2}{m_{f\bar{f}}^2}\frac{m_{f\gamma}^2}{m_{\bar{f}\gamma}^2}}\;
\right)\,, \\
\beta_{W\pm} & = &\case{1}{2}\left(1 \pm \sqrt{1 - 4m_W^2/m_H^2}\;\right)\, \\
\gamma_{\pm} & = &\case{1}{2}\left(1 \pm \sqrt{1 - 4m_W^2/m_{f\bar{f}}^2}\;
\right)\,.
\end{eqnarray}

The various $C_0$'s are
\begin{eqnarray}
C_0(1,2,3) & = & \frac{1}{m_{f\bar{f}}^2}\Biggl\{
- {\rm Sp}\left(\frac{m_W^2}{m_W^2 - m_{f\bar{f}}^2\,\gamma_+}\right)
+ {\rm Sp}\left(\frac{m_W^2 - m_{f\bar{f}}^2}{m_W^2 - m_{f\bar{f}}^2\,
\gamma_+}\right) \nonumber \\
&   &
- {\rm Sp}\left(\frac{m_W^2}{m_W^2 - m_{f\bar{f}}^2\,\gamma_-}\right)
+ {\rm Sp}\left(\frac{m_W^2 - m_{f\bar{f}}^2}{m_W^2 - m_{f\bar{f}}^2\,
\gamma_-}\right)
-{\rm Sp}\left(\frac{m_W^2 - m_{f\bar{f}}^2}{m_W^2}\right)\nonumber \\
&   &
+ \frac{\pi^2}{6}
+ i\pi\Biggl[\theta(m_{f\bar{f}}^2 - 4m_W^2)\ln\left|\frac{m_W^2 -
m_{f\bar{f}}^2\,\gamma_+}{m_W^2 - m_{f\bar{f}}^2\,\gamma_-}\right|\;
\Biggr]\;\Biggr\}\,, \\ [8pt]
C_0(1,2,4) & = & \frac{-1}{m_{\bar{f}\gamma}^2 - m_H^2}\Biggl\{
{\rm Sp}\left(\frac{\lambda_0}{\lambda_0 - \lambda_{1}}\right)
- {\rm Sp}\left(\frac{\lambda_0 - 1}{\lambda_0 - \lambda_{1}}\right)
\nonumber \\
&   &
- {\rm Sp}\left(\frac{\lambda_0}{\lambda_0 - \beta_{W+}}\right)
+ {\rm Sp}\left(\frac{\lambda_0 - 1}{\lambda_0 - \beta_{W+}}\right)
- {\rm Sp}\left(\frac{\lambda_0}{\lambda_0 - \beta_{W-}}\right)
\nonumber \\
&   &
+ {\rm Sp}\left(\frac{\lambda_0 - 1}{\lambda_0 - \beta_{W-}}\right)
- {\rm Sp}\left(\frac{\lambda_0 - 1}{\lambda_0}\right)
+ \frac{\pi^2}{6} \nonumber \\
&   &
+ i\pi\Biggl[\theta(m_H^2 - 4m_W^2)\ln\left|\frac{\beta_{W+} - \lambda_0}
{\beta_{W-} - \lambda_0}\right|
- \theta(m_{\bar{f}\gamma}^2 - m_W^2)\ln\left|\frac{\lambda_1 - \lambda_0}
{-\lambda_0}\right|\;\Biggr]\;\Biggr\}\,, \\ [8pt]
C_0(1,3,4) & = & \frac{-1}{m_{f\bar{f}}^2 - m_H^2}\Biggl\{
{\rm Sp}\left(\frac{1}{\beta_{W+}}\right) 
+ {\rm Sp}\left(\frac{1}{\beta_{W-}}\right)
- {\rm Sp}\left(\frac{1}{\gamma_+}\right) 
- {\rm Sp}\left(\frac{1}{\gamma_-}\right) \nonumber \\
&   &
+ i\pi\Biggl[\theta(m_H^2 - 4m_W^2)\ln\left|\frac{\beta_{W+}}{\beta_{W-}}
\right| 
- \theta(m_{f\bar{f}}^2 - 4m_W^2)\ln\left|\frac{\gamma_+}{\gamma_-}\right| 
\;\Biggr]\;\Biggr\}\,, \\ [8pt]
C_0(2,3,4) & = & \frac{1}{m_{\bar{f}\gamma}^2}\Biggl\{
{\rm Sp}\left(\frac{m_{\bar{f}\gamma}^2}{m_W^2}\right)
+ i\pi\theta(m_{\bar{f}\gamma}^2 - m_W^2)
\ln\left|\frac{m_{\bar{f}\gamma}^2}{m_W^2}\right|\;\Biggr\}\,,
\end{eqnarray}
with
\begin{eqnarray}
\lambda_0 & = & \frac{m_W^2}{m_H^2 - m_{\bar{f}\gamma}^2}\,,\\
\lambda_1 & = & \frac{m_{\bar{f}\gamma}^2 - m_W^2}{m_{\bar{f}\gamma}^2}\,.
\end{eqnarray}

In this case, the $B_0$'s are
\begin{eqnarray}
B_0(1,3) & = &\;\Delta + 2
+ \gamma_+\ln\left(\frac{\gamma_+ - 1 + i\varepsilon}{\gamma_+}\right) 
+ \gamma_-\ln\left(\frac{\gamma_- - 1 - i\varepsilon}{\gamma_-}\right) 
\,, \\ [8pt]
B_0(1,4) & = &\;\Delta + 2
+ \beta_{W+}\ln\left(\frac{\beta_{W+} - 1 + i\varepsilon}{\beta_{W+}}\right) 
+ \beta_{W-}\ln\left(\frac{\beta_{W-} - 1 - i\varepsilon}{\beta_{W-}}\right)
\,, \\ [8pt]
B_0(2,4) & = &\;\Delta + 2 
- \left(1 - \frac{m_W^2}{m_{\bar{f}\gamma}^2}\right)\Biggl[\ln\left|
1 - \frac{m_{\bar{f}\gamma}^2}{m_W^2}\right|
- i\pi\theta(m_{\bar{f}\gamma}^2 - m_W^2)\,\Biggr]\,,
\end{eqnarray}
and 
\begin{equation}
\Delta = \pi^{(n/2 -2)}\left(\frac{m_W^2}{\mu^2}\right)^{(n/2 -2)}
\Gamma\left(2 - \frac{n}{2}\right)\,.
\end{equation}

\newpage


\begin{references}
\bibitem{tri} There are also diagrams with two gauge boson poles, but these
actually vanish in the gauges we use for the perturbative calculation.
\end{references}
\end{document}